\documentclass{elsart}

 \def\agt{\mathrel{\hbox{\rlap{\hbox{\lower4pt\hbox{$\sim$}}}\hbox{$>$}}}}
 \def\alt{\mathrel{\hbox{\rlap{\hbox{\lower4pt\hbox{$\sim$}}}\hbox{$<$}}}}
 \usepackage{amssymb}
 \usepackage{bm}

 \usepackage[dvips]{graphics}
\usepackage{amssymb}

\begin{document}

\begin{frontmatter}



\title{Efficient method of finding scaling exponents from finite-size Monte-Carlo simulations}


\author{Jaan Kalda}

\address{CENS, Institute of Cybernetics,
Tallinn University of Technology,
Akadeemia tee 21,
12618 Tallinn,
Estonia}

\begin{abstract}
Monte-Carlo simulations are routinely used for estimating the scaling exponents of complex systems. 
However, due to finite-size effects, determining the exponent values is often difficult and not reliable.
Here we present a novel technique of dealing the problem of finite-size scaling. The efficiency of the technique is 
demonstrated on two data sets.
\end{abstract}

\end{frontmatter}

% main text
%\section{}
%\label{}

% \appendix

% \section{}
% \label{}
\section{Introduction}
Determining the scaling exponents from the finite-size simulation data 
is a very common task in the physics of complex systems.
In particular, this technique is widely used in the context of phase transitions, surface roughening, turbulence, granular media, etc, c.f. reviews \cite{Barber,Binder,Privman}.
Typically, the finite size Monte-Carlo studies involve  extrapolation of the simulation data towards infinity. 
Unless there is some theoretical understanding about the functional form of the finite size corrections to the asymptotic scaling laws of the particular system, 
such an extrapolation carries a risk of underestimating the uncertainties.
In some cases, it may be helpful to  increase the computation time and system size, and optimise the  simulation scheme (c.f. \cite{JKMC}).
However, this is not always feasible, because the convergence to the asymptotic scaling law may be very slow. 
Additional difficulties arise, when one needs to determine the exponents of the finite-size correction terms %to the asymptotic law 
(c.f.\ \cite{Bischof}), or when the asymptotic power law includes a logarithmic pre-factor.

%	(e.g. in a logarithmic form \cite{Bischof} or integer-power \cite{Salas,Oliveira}	
In what follows, we develop a novel technique of determining scaling exponents from the finite size simulation data.
Using two different simulation series as examples, we demonstrate its effectiveness. The scaling exponents are found with high 
accuracy,  even in the case of  an extremely slow asymptotic convergence (when a straightforward power law fit is effectively unusable). 
Furthermore, we are able to calculate the exponents of the first and second correction terms to the asymptotic scaling law.

\section{Description of the method}
Let us consider a model system, described by its size $N$, assuming that 
the smallest possible value of $N$ plays the role of the unit length.
Suppose  that  the mathematical expectation 
of a certain physical quantity scales as $L\propto N^\alpha$ for $N\gg 1$.
For illustration, if our model system is a percolation lattice inside a square of side length $N$ at criticality, then the quantity $L$ 
could be the mass of its largest percolation cluster.
The  Monte-Carlo simulations can be used to estimate the values of the mathematical expectation $L$ for several system sizes 
\[N_1 < N_2 < \ldots < N_n;\]
let us denote these estimates as ${\cal L}_i$, $i=1,\ldots n$,
and the variances of them as $\sigma_i^2$. Then,
root-mean-square fit can be used to obtain the scaling exponent $\alpha$, c.f.\ \cite{Binder}. 
However, it is often difficult to estimate the uncertainty of the obtained result, because the magnitude of the finite-size corrections to the asymptotic law $L=L_0N^{\alpha_0}$ is unknown.
Of course, one can plot $\ln {\cal L}_i$ versus $\ln N_i$ and determine such a transition point $i=k$ that for $i\ge k$, the data points lay within their statistical uncertainties on a straight line.
Then, only the data points with $i\ge k$ will be used for finding the exponent $\alpha_1$. However, one can easily underestimate the adequate value of $k$, because
the statistical fluctuations just happen to compensate the finite-size corrections $\Delta L$. On the other hand, taking excessively large values of $k$ would inflate  the 
variance  $\sigma_{\alpha_1}$ of the outcome. Finally, in some cases, the decay rate of the corrections $\Delta L$ can be very slow (see below), so that the method outlined above will
fail at the first step --- there is no linear range of the graph.

This problem can be resolved, if it is possible to find  more than one physical quantity with similar scaling behaviour.
Suppose one can define $m$ distinct quantities, the mathematical expectations $L_k$ ($k=1, 2, \ldots m$) of which obey identical scaling exponents (examples will be provided in the next Section).
%In what follows, we use $m=3$; however, the method is easily generelized to larger values (and also to $m=2$).
The method will work, if the following additional conditions are satisfied. First, 
the mathematical expectations  $L_k$ can be expanded asymptotically as
%This can be explained in terms of asymptotic expansion
%Consequently, if we assume that $l_i$ can be expanded asymptotically,
\begin{equation} \label{Lsum}
	L_k(N)=\sum_{\mu=1}^\infty A_{\mu k} N^{\alpha_{\mu k}},\;\;\;\mbox{with}\;\;\; \alpha_{(\mu+1)k} < \alpha_{\mu k};
\end{equation}
second, the first $m$ leading exponents $\alpha_{\mu k}, \mu=1,\ldots m$ are equal, so that we can designate
\begin{equation}
	\alpha_{\mu}\equiv \alpha_{\mu k},\;\;\; \mbox{for}\;\;\; \mu \le m.
\end{equation}
Third,  the first $m$ leading terms in the expansion (\ref{Lsum})  are linearly independent, $\det A_{\mu k} \ne 0$ (with $\mu, k \le m$).
We shall find out later, how to verify, if these conditions are satisfied.
%the non-leading exponents ($\mu \ne 1$) are close to the leading term. 

Now, let us designate the sum of the residual terms as 
\[\delta_i(N)\equiv\sum_{\mu=m+1}^\infty A_{\mu i} N^{\alpha_{\mu i}},\]
and introduce $B_{\mu k}$ as the 
$m\times m$ inverse matrix of $ A_{\mu k}$.
Then, the relationship 
\[
	\sum_{\mu=1}^m A_{\mu k} N^{\alpha_\mu }=L_i(N)-\delta_k(N)
\]
can be rewritten as 
\begin{equation} \label{Lsum}
	 N^{\alpha_\mu}=\sum_{k=1}^m L_k(N)B_{k\mu}-\Delta_\mu(N),
\end{equation}
where 
\begin{equation} \label{Delta}
\Delta_\mu(N)=\sum_{k=1}^m \delta_k(N)B_{k\mu}
\end{equation}
So, neglecting the residual term $\Delta_\mu(N)$, the power law $N^{\alpha_\mu}$ can be expressed as a linear combination of the 
$m$ functions $L_k(N)$. Hence, if we perform a least-square fit of the $n$-dimensional vector $\vec x^d\equiv (N_1^d, N_2^d,  \ldots N_n^d)$ with a linear combination of the $m$
data vectors $\vec {\cal L}_k \equiv ({\cal L}_{k1}, {\cal L}_{k2}, \ldots {\cal L}_{kn})$, and plot the sum of the squared residuals $S(d)$  as a 
function of $d$, then there should be $m$ minima, at $d=\alpha_1, \alpha_2, \ldots \alpha_m$.
Here, ${\cal L}_{ki}$ denotes the Monte-Carlo estimate of the expectation $L_k(N)$ at $N=N_i$.
Indeed, according to Eq.~(\ref{Lsum}), such a fit should be possible for the listed values of $d$ [assuming that the statistical fluctuations dominate over the residual terms $\Delta_\mu(N)$].
So, if the function $S(d)$ does, indeed, have $m$ clear minima, then the validity of the above mentioned assumptions is confirmed, and the 
positions of the minima can be used to find the values of the exponents $\alpha_\mu$ for $\mu=1, \ldots m$.

Let us consider this method in more details. So, we need to find the function
\begin{equation} \label{sumR}
	S(d)=\sum_{i=1}^n\left(N_i^d-\sum_{k=1}^m C_k{\cal L}_{ki}\right)^2s_i^{-2},
\end{equation}
where the constants $C_k$ are optimised, yielding the minimal value of the expression in right-hand-side; here $s_i^2$ denotes the variance of the expression between the braces, 
\[
s_i^2=\sum_{k,l=1}^n C_kC_l\Sigma_{kli},
\]
and $\Sigma_{kli}$ is the covariance matrix of the Monte-Carlo simulation results for $L_k(N)$ at $N=N_i$.
The covariance matrix is needed, because in order to save the computation time, it is reasonable to use the same simulation data for 
all the quantities $L_k$. These quantities are probably strongly correlated, because they describe similar aspects of the model system.
So, a proper statistical analysis requires the covariance matrix. Fortunately, it can be easily estimated using the same Monte-Carlo simulation data, without noticeable 
increase in computing resources.

If the weighting factors $s_i^{-2}$ in Eq.~(\ref{sumR}) were constant, then the problem of finding the function $S(d)$ would be a simple linear least-square fitting task. 
Things are slightly more complicated due to  the fact that the weighting factors depend on the constants $C_k$. However, if the variances $s_i^{2}$ are calculated iteratively
(using the constants $C_k$ from the previous iteration), then the convergence is very fast (typically, no more than three iterations are required).

Notice that  at the optima, i.e.  for $d=\alpha_\mu$ ($\mu< m$), the random variable $S(d)$ should have chi-square distribution with $n-m-1$ degrees of freedom,
if all the three above mentioned assumptions are fully satisfied. So, the values at the minima, $S(\alpha_\mu)$, can be compared with the critical values $\chi_{n-m-1}^2(p)$ of the
chi-square distribution, to further  examine the validity of these assumptions. If the result is positive [i.e. $S(\alpha_k) < \chi_{n-m-1}^2(p)$], the same critical values can be used
to estimate the uncertainties of the results by finding such values $\Delta_{\alpha_k}$ that $S(\alpha_k\pm \Delta_{\alpha_k}) \approx \chi_{n-m-1}^2(p)$.

As a final remark, let us notice that one could use the power series (\ref{Lsum}) for a direct nonlinear least-square fitting. However, nonlinear fitting by itself is a difficult task, if there is
a large number of fitting parameters. What is more important, a larger number of  fitting variables would be needed, to take into account the same number of terms in the asymptotic expansion 
(i.e.\ to achieve the same accuracy). Indeed, here, we have $m+1$ fitting parameters
($C_k$ and $d$). A straightforward fit with $m$ first terms in the expansion  (\ref{Lsum}) would include $2m$ fitting parameters ($A_{\mu k}$ and $\alpha_\mu$,  $\mu=1,\ldots m$). 
Out of those $2m$ parameters, half are nonlinear ones ($\alpha_\mu$). As a result, it would be practically impossible to handle more than two terms in the power series.
Besides, more fitting parameters  results in larger uncertainties of the fitting results.

\section{First example: statistical topography of rough surfaces}
Here we consider the simulations, which were performed to determine the fractal dimension of a 
certain set of contour lines of random Gaussian self-affine surfaces. This set of contour lines will be referred to as the 
``oceanic coastline''. Providing detailed discussions, why it was necessary to define and study the ``oceanic coastlines'', is beyond the scope of the present paper.
In what follows, only as much details will be provided, as is needed for illustrating our new technique of determining the scaling exponents.

Let us consider  a random surface, which is given by the surface height $\psi(x,y) \equiv \psi(\bm{r})$ 
over a two-dimensional plane. It is assumed that this surface is Gaussian and self-affine, 
characterized by  the Hurst exponent $H$:
\begin{equation}
	\left< [\psi(\bm{r})-\psi(\bm{r}+\bm{a}) ]^2\right>  \propto |\bm{a}|^{2H}.
\end{equation}
Here, the angular braces denote averaging over different realizations of the surface; 
we assume that $0\le H\le 1$.

Further, let the surface be flooded by water up to a level $h$.
Then, regions with $\psi(\bm{r}) < h$  will be called ``wet''.
Now, we pick a connected (possibly infinite)  wet
region and name it ``ocean'' (all the other connected wet regions are called ``lakes'').
The perimeter of the ``ocean'' is called the ``oceanic coastline'' 

The fractal dimension of the ``oceanic coastline'' has been calculated numerically,
using the following method.
Instead of isotropic two-dimensional surfaces, 1+1-dimensional [(1+1)D] random surfaces 
are generated using the lattice of the four-vertex model \cite{4VM}. Such surfaces are 
assumed to belong to the same universality class as the statistically
isotropic 2D surfaces, but are numerically more efficient. 
So, the surface height is given by
\[
\psi(x,y)= f_{H}(\lfloor x \rfloor)-g_{H}(\lfloor y \rfloor),
\]
where $f_{H}$ and $g_{H}$ are two uncorrelated one-dimensional discretised fractional Brownian functions, 
which take only integer values and satisfy additional constraint
\begin{equation}\label{fh}
f_H(j) = f_H(j-1) \pm 1,
\end{equation}
Here, $\lfloor x \rfloor$ denotes the floor function of $x$. 
These one-dimensional functions are generated using
uncorrelated random sequences of ``spins'' $s_i=\pm 1$ ($i \in  \mathbb{N}$).
Each ``spin'' sequence defines an aim function
\begin{equation}
	F_H(j) = \sum_{i=0}^{j-1} s_j |j-i|^{H-0.5}, 
\end{equation}
where $i,j \in \mathbb{N}$. The aim function  is approximated by a function $f_H(j)$, as closely as possible under the constraint (\ref{fh}). 

The simulations have been performed for square polygons of side length 
$N_1=128$, $N_2=196$, $N_3=256$, $N_4=392$, \ldots $N_9=2048$, using different values of $H$.
For each realization of the surface, such an ``oceanic coastline''
has been found, which connects a pair of opposite edges of the polygon.
% has been found by simulating a flooding process
This has been achieved by simulating the following surface flooding process. The polygon boundary is assumed to be impenetrable for the water.
The water is  injected slowly onto the surface at the lowest point of the perimeter of the polygon.
Water injection is terminated as soon as the flooded region 
connects a pair of opposite edges of the polygon. 
At the end of such a flooding, three quantities were
recorded: $L_1$ --- the coastline length, $L_2$ --- the number of cells (i.e. the faces of the square lattice) touching the coastline, and $L_3$ --- the coastline length 
immediately before achieving the critical flood level. Apparently, all these quantities have the same asymptotic scaling exponent $\alpha_1$, i.e.  $L_k \propto N^{\alpha_1}$ with $k=1,2$ and $3$. 
For each polygon size and Hurst exponent value, 
$M=10^8$ or more surfaces have been generated.  

\begin {figure}[htb]
\includegraphics{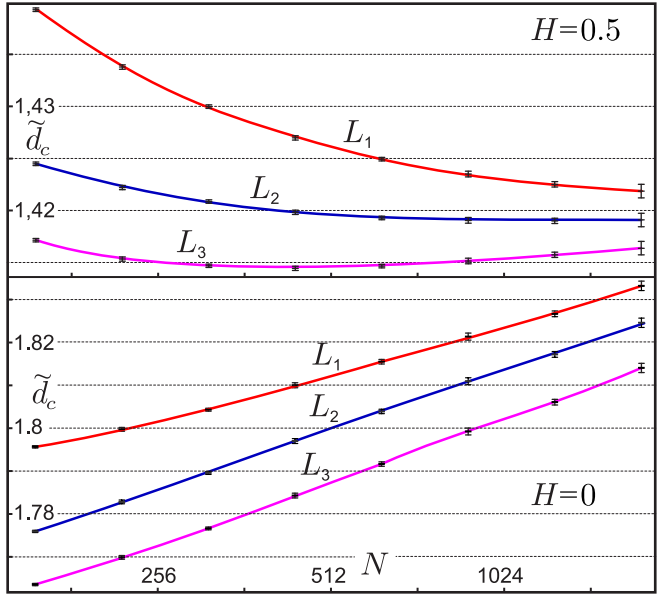}
\caption{The values of the differential fractal dimension of the ``oceanic coastline'' $\tilde d_c=\log_2 ({\cal L}_{ki} /{\cal L}_{ki+1})$ are plotted versus the system size $\sqrt{N_iN_{i+1}}$ in semilogarithmic graph. 
Upper three curves correspond to the Hurst exponent $H=0.5$, lower curves --- to $H=0$. While upper curves seem to converge around $\tilde d_c\approx 1.419$, no convergence can be observed for $H=0$.
Using the same simulation data, our new method yields $\tilde d_c\approx 1.4203\pm 0.0009$ for $H=0.5$ and $\tilde d_c\approx 1.8975\pm 0.0025$  for $H=0$.}
\end {figure}

\begin {figure}[htb]
\includegraphics{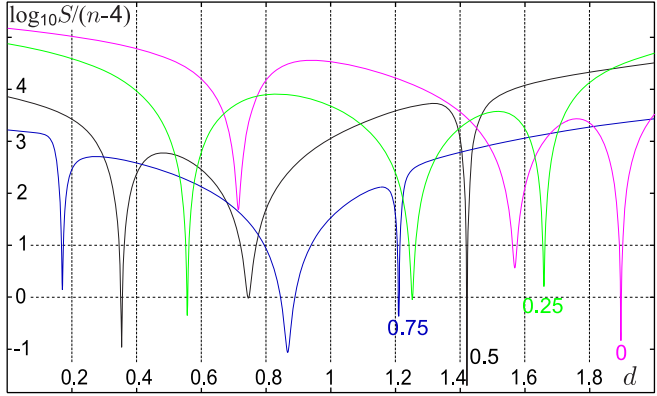}
\caption{The logarithm of the sum of squared residuals (reduced to the number of degrees of freedom), $\log_{10} [S(d)/(n-4)]$, is plotted versus $d$ for different 
values of the Hurst exponent $H$ (the curves are labeled with numbers indicating the value of $H$). 
For all the curves, three deep minima are present. The positions of these minima allow us to determine the exponents of the asymptotic expansion [Eq.~(\ref{Lsum})].}
\end {figure}

In Fig,~1, the strength of finite size effects is demonstrated by plotting the differential fractal dimension, defined as  $\tilde d_c=\log_2 ({\cal L}_{ki} /{\cal L}_{ki+1})$, versus the system size $\sqrt{N_iN_{i+1}}$.
For $H=0.5$, these curves seem to suggest that the asymptotic value of the oceanic coastline fractal dimension is $d_c=1.419\pm 0.002$. For $H=0$, the finite size effects are so strong that it is 
impossible to give any estimate for $d_c$. In Fig.~2,
the curves $\log_{10} [S(\alpha)/(n-4)]$ are plotted versus the exponent value $d$. 
The existence of three sharp minima for all the curves confirms the validity of the assumptions required for the applicability of our new method. In particular, it is possible to conclude that  for 
$H=0.5$, $\tilde d_c\approx 1.4203\pm 0.0009$ and for $H=0$, $\tilde d_c\approx 1.8975\pm 0.0025$. Also, it is possible to determine the values of the second and third exponents of the asymptotic expansion.
For $H=0.5$, these values are $\alpha_2=0.745\pm 0.015$ and $\alpha_3=0.353\pm 0.004$; for $H=0$, $\alpha_2=1.563\pm 0.005$ and $\alpha_3=0.721\pm 0.005$.
The uncertainties here have been obtained using the critical coefficients for the chi-square distribution for $p=95\%$. 

Finally, let us study the shape of the curve for $H=0$ in Fig.~2 in more details. This curve corresponds to $n-4=4$ degrees of freedom. The last minimum (at $d=\alpha_1$) is so deep that the
residual term $\Delta_\mu$ in Eq.~(\ref{Lsum}) is clearly negligible. However, the minima at $d=\alpha_2$ and $\alpha_3$ do not pass the test based on the  chi-square distribution.
The difference in depth  of the minima is explained by two circumstances. First, the residual terms $\Delta_\mu$ can be of different amplitude for different values of $\mu$, because
the terns in Eq.~(\ref{Delta}) can efficiently cancel out (likewise, they can also magnify each other). Second, the variance of the linear combination in Eq.~(\ref{Lsum}) can also depend considerably 
on $\mu$, due to the non-diagonal elements of the covariance matrix $\Sigma_{kl}$. Therefore, in order to determine the values  of $\alpha_2$ and $\alpha_3$, it was necessary to 
skip respectively one and two datapoints (at $N=N_1$ and $N=N_2$), reducing the number of degrees of freedom down to 3 or 2.

%Note that for smaller cut-off values ($N_0=128$ or $196$), the residual term  $o(M^{\alpha_2})$ was no longer 
%negligible, resulting in 
%the minima were unacceptably large [with $S/(n-4) \gtrsim 1$] is comparable
%to the variance $\sigma_M$, and leads to unacceptably large minima [with $S/(n-4) \gtrsim 1$].

\section{Second example: hulls of 1+1-dimensional percolation problem}
%\begin {figure}[!b]
%\includegraphics{fig5.eps}
%\caption{The mismatch between a linear estimate $d_{lin}=\frac{91}{48}-H\frac{43}{48}$ and the fractal dimension of 
%the ``oceanic coastline'' $d_c$ is plotted versus the Hurst exponent $H$.} 
%\end {figure}

%FIG5
%

%FIG6
\sloppy
Here we consider a simple modification of the two-dimensional percolation problem, when the bond breaking process is defined by two functions of one variable (hence, we call it 1+1 dimensional).
More specifically, we use the surrounding lattice, the sites of which are at the middlepoints of the of the percolation lattice. 
Let the $x$ and $y$ axes be defined so that the sites of the surrounding lattice have integer coordinates. Further, if we have two 
functions $f(x)$ and $g(y)$, which take random uncorrelated values $+1$ or $-1$ for each integer $x$ and $y$, then the
bond of the percolation lattice with middlepoint at $(x,y)$ is broken, if $f(x)g(y)=-1$; the bond is present, if $f(x)g(y)=1$. 

We study the scaling of  the length of the hull of a percolation cluster as a function of its diameter. To this end, we use Monte-Carlo simulations to estimate mathematical expectations of three quantities 
for such hull segments, which starts at the origin, and reach the edge of  a square polygon of side length $2N$.
These quantities are defined as follows:
 \begin{description}
\item{ $L_1$} --- the length of the hull, measured in the number of surrounding lattice elements;\\
\item {$L_2$} --- the number of such segments of the hull, which have {\em three consecutive clockwise turns} or three consecutive 
counter-clockwise turns (and hence, consist of four elements of the surrounding lattice); \\
 \item {$L_3$} ---  the number of such segments of the hull, which have {\em four alternating turns} 
(left-right-left-right or right-left-right-left). 
\end{description}

The simulation results are again illustrated by two figures. In Fig,~3, the differential fractal dimension [defined as before,  $\tilde d_c=\log_2 ({\cal L}_{ki} /{\cal L}_{ki+1})$], 
is plotted versus the polygon size $\sqrt{N_iN_{i+1}}$.
\begin {figure}[htb]
\includegraphics{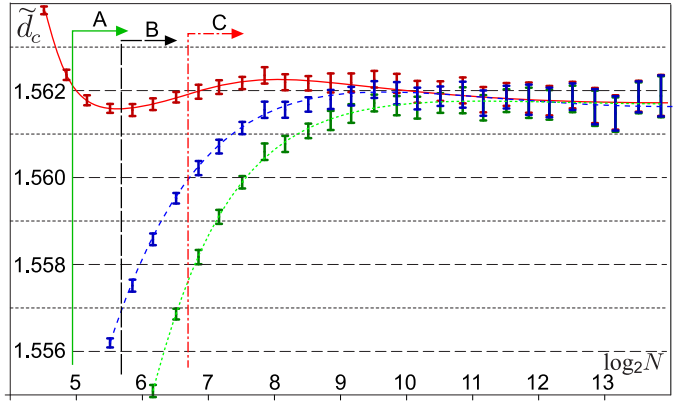}
\caption{The differential fractal dimension of the hull of the (1+1)D percolation clusters  $\tilde d_c=\log_2 ({\cal L}_{ki} /{\cal L}_{ki+1})$ is plotted 
versus the system size $\sqrt{N_iN_{i+1}}$ in semilogarithmic graph, similarly to Fig.~1.
}
\end {figure}

Unlike in the case of ``oceanic coastlines'', here the polygon size can be dynamically enlarged: we track a hull, and at the moment when it reaches the polygon 
boundaries at $N=N_i$, we record the data for $N_i$ and enlarge the polygon size up to $N=N_{i+1}$. We start with a new hull, if the hull forms a closed loop, or if $N_i$ reaches
the maximal allowable value. The benefit is that we can make the array of datapoints more dense, without spending additional computing time. The drawback is that
there will be correlations between the neighbouring datapoints, hence the statistical analysis will become more complicated.

In order to take into account the correlations  in a correct way, it would be necessary to perform a linear transform of the $n$-dimensional space of data-vectors, making the covariance matrix diagonal.
Then, ordinary least-square fit could be performed in the new system of coordinates.
However, as a simplified approach, one can calculate still the sum of squared residuals $S(d)$,  initially  ignoring the correlations between the data points.
By doing so, we keep the minima of the function $S(d)$ in their correct places, i.e.\ the estimates for the scaling exponent values remain completely correct.
However, our error analysis will be approximate, because 
the distribution of $S(\alpha_\mu)$ will no longer be  the chi-square one with $n-m$ degrees of freedom. The effective number of  degrees of freedom
will be somewhat smaller: $n$ needs to be substituted by an efficient number of uncorrelated datapoints $n_{\mbox{\scriptsize eff}}<n$. The correlated data fluctuations contribute to the fluctuations of  
$S(d)$ by enhancing each other. As a result, $S(d)$ will have a chi-square distribution with $n_{\mbox{\scriptsize eff}}-m$ degrees of freedom, scaled by a factor of $\kappa=(n/n_{\mbox{\scriptsize eff}})^2$.
The covariance matrix can be used to estimate the value of  $n_{\mbox{\scriptsize eff}}$.

\begin {figure}[htb]
\includegraphics{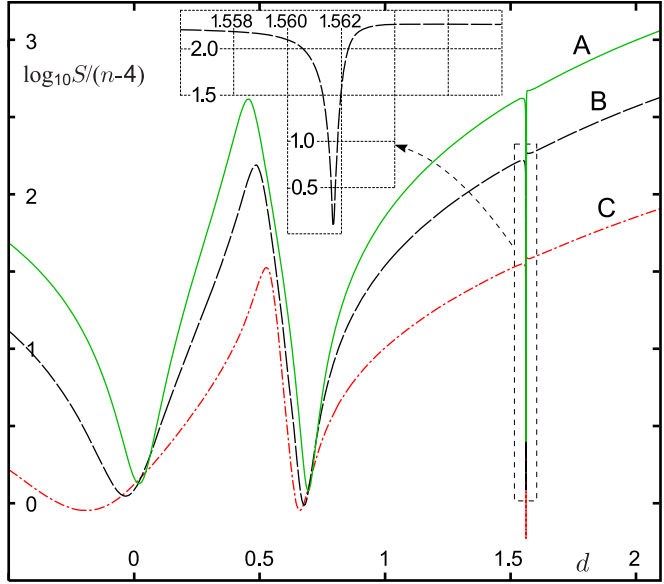}
\caption{Using the same simulation data as in Fig.~3, the logarithm of the sum of squared residuals is plotted versus $d$ for different 
values of datapoints (the respective datapoint ranges are also marked in Fig 3.). 
For all the curves, three deep minima are present.}
\end {figure}
This approach has been used in Fig.~3, with $n/n_{\mbox{\scriptsize eff}}=3$, where $\log_{10} [\kappa^{-1}S(d)/(n_{\mbox{\scriptsize eff}}-4)]$ is plotted versus the exponent $d$, for different values of $n$. 
Different curves correspond to different number of skipped data points (the starting points of the respective data ranges 
are indicated in Fig.~3 by vertical lines $A$, $B$, and $C$). Insert provides a zoomed region around the sharp minimum for the curve labeled by $B$.
These results allow us to conclude that $\alpha_1 = 1.56166\pm 0.00008$, which is consistent with Fig.~3, but with increased precision.
Also, we can conclude that $\alpha_2=0.66\pm 0.03$, and $\alpha_3=-0.07\pm 0.14$. It is most likely that in fact, $\alpha_3=0$. Indeed, 
$\alpha_3=0$ corresponds to a constant offset of quantities $L_k$, which would appear immediately, if (for instance) $L_1$ is measured without the first element (of the surrounding lattice).

\section{Conclusions}
A novel and universal method of determining the scaling exponents via finite-size Monte-Carlo simulations has been devised.
The method can be applied, if it is possible to find $m\ge 2$ distinct quantities with equal scaling exponents. 
The two above considered  examples provide general guidelines, how to define such quantities, and suggest that typically, these quantities can be indeed found.
Here, we have used $m=3$, which provides a  good cancellation of higher order corrections to the asymptotic scaling law, and keeps the number of least-square fitting parameters reasonably small.

For all the considered cases, the method increases the accuracy of the scaling exponent estimates. However, the method is particularly useful in the case of large corrections to the asymptotic scaling law
[such as demonstrated in Fig.~1 ($H=0$)]. Also, the method is extremely useful, if it is necessary to find the exponents of the finite size correction terms.

The support of Estonian Science F	oundation grant No 6121 is acknowledged.

\bibliography{surf}{}
\bibliographystyle{elsart-num}

%\begin{thebibliography}{00}


% notes:

% \bibitem{label2}
% If there is a note, it should come last:
% \bibitem{label3} \note
% \end{subbibitems}

%\bibitem{}

%\end{thebibliography}

\end{document}